\documentclass[conference,review=true,anonymous=true]{IEEEtran} 
\IEEEoverridecommandlockouts

\usepackage{etoolbox}
\usepackage{ifthen}

\newboolean{makeXTRA}
\setboolean{makeXTRA}{false} 

\makeatletter%
\newboolean{maketitleAfterAllMeta}
\@ifclassloaded{acmart}%
{\setboolean{maketitleAfterAllMeta}{true}}%
{\setboolean{maketitleAfterAllMeta}{false}}%
\makeatother%

\usepackage[utf8]{inputenc}
\usepackage[T1]{fontenc}

\usepackage{microtype}

\usepackage{anyfontsize}
\usepackage[10pt]{moresize}

\usepackage{amsmath}
\usepackage{amsfonts}
\usepackage{mathtools}

\makeatletter%
\@ifclassloaded{acmart}{}{\usepackage{amssymb}}
\makeatother%

\usepackage{siunitx}
\usepackage{xfrac}
\usepackage{tabu}
\usepackage{xspace}

\usepackage[normalem]{ulem}

\makeatletter%
\@ifclassloaded{acmart}{}%
{\usepackage[usenames,dvipsnames,table]{xcolor}}
\makeatother%

\makeatletter%
\@ifclassloaded{acmart}{}
{\@ifclassloaded{elsarticle}{}
{\usepackage{cite}}}
\makeatother%

\usepackage{setspace}

\usepackage{verbatim}

\usepackage{booktabs}
\usepackage{makecell}
\usepackage{multicol}
\usepackage{multirow}
\usepackage{tabularx}
\usepackage{array}

\usepackage{stfloats}

\makeatletter
\@ifclassloaded{IEEEtran}{%
  \let\MYcaption\@makecaption%
  \usepackage[font=footnotesize]{subcaption}%
  \let\@makecaption\MYcaption%
}{%
  \usepackage{subcaption}%
}%
\makeatother

\usepackage[textsize=tiny]{todonotes}

\makeatletter%
\@ifclassloaded{IEEEtran}{}{}%
\usepackage{enumitem}
\makeatother%

\usepackage{quoting}

\usepackage{listings}

\lstdefinestyle{mystyle}{
    language=Python,           
    basicstyle=\ttfamily,      
    keywordstyle=\color{blue}, 
    frame=single,              
    breaklines=true,           
    captionpos=b               
}

\makeatletter%
\@ifclassloaded{acmart}%
{}{}%
\@ifclassloaded{elsarticle}%
{}{}
\@ifclassloaded{llncs}%
{}{}
\@ifclassloaded{jpconf}%
{%
  \usepackage{doi}%
}{}
\@ifclassloaded{IEEEtran}%
{%
  
  \newcommand{\ifIEEEtran}[1]{#1}
}%
{%
  \newcommand{\ifIEEEtran}[1]{}
  \usepackage{flushend}
}%
\makeatother%

\makeatletter%
\@ifpackageloaded{hyperref}%
{}%
{%
  \usepackage[hyphens]{url}
  \usepackage[breaklinks=true,bookmarks=false]{hyperref}
  \hypersetup{
    colorlinks=false,%
    pdfborder = 0 0 0
  }
}%
\makeatother%

\usepackage[capitalize,noabbrev,capitalize]{cleveref} 

\usepackage{stackengine}
\usepackage{graphicx}
\usepackage{tikz}
\usepackage{pgfplots}
\usepackage{pgfplotstable}

\ifthenelse{\boolean{makeXTRA}}
  {\newcommand{\XTRA}[1]{\phantom{}\begingroup\slshape\color{RoyalBlue}\ignorespaces#1\ignorespaces\endgroup}}
  {\newcommand{\XTRA}[1]{}}

\DeclareMathAlphabet{\mathpzc}{OT1}{pzc}{m}{it}

\makeatletter
\newcommand*{\textoverline}[1]{$\overline{\hbox{#1}}\m@th$}
\makeatother

\makeatletter
\@ifclassloaded{IEEEtran}{%
}{%
  
}%

\makeatletter 
\newcommand{\linebreakand}{%
  \end{@IEEEauthorhalign}
  \hfill\mbox{}\par
  \mbox{}\hfill\begin{@IEEEauthorhalign}
}
\makeatother 

\newlist{myReuseInterval}{enumerate}{1}
\setlist[myReuseInterval]{label=(R$_{\arabic*}$),nosep}

\crefname{section}{\S}{\S}
\crefformat{section}{#2\S#1#3}

\crefname{figure}{Fig.}{Figs.}
\Crefname{figure}{Figure}{Figures}

\crefname{equation}{Eq.}{Eqs.}

\definecolor{shadecolor}{gray}{0.9} 

\makeatletter
\@ifpackageloaded{algorithm2e}{\@ifclassloaded{IEEEtran}{%
  \SetAlFnt{\small}
  \SetAlCapFnt{\small}
  \SetAlCapNameFnt{\small}
}}%
\makeatother

\newcommand{\lstsetCommon}{%
  \lstset{%
    columns=fullflexible,%
    keepspaces=true,%
    escapeinside={<[}{]>},%
    moredelim=**[is][\color{Cerulean}]{<*}{*>},
    moredelim=**[is][\color{Red}]{<^}{^>},
    basicstyle=\ttfamily\footnotesize,%
    showstringspaces=false,%
    aboveskip=0em,%
    belowskip=0em,%
  }%
}

\newcommand{\lstsetMe}{%
  \lstsetCommon{}%
  \lstset{%
  }
}

\lstsetMe{}

\usepackage{soul}

\newif\ifhl{}
\hltrue{}
\hlfalse{}

\ifhl{}
 
\else
 
\fi

\newif\ifdraft{}
\drafttrue{}

\ifdraft{}
  \newcommand{\davidnote}[1]{ {\textcolor{purple} { ***[David]: #1 }}}
  \newcommand{\raynote}[1]{ {\textcolor{orange} { ***[Ray]: #1 }}}
  \newcommand{\ozgurnote}[1]{ {\textcolor{blue} { ***[Ozgur]: #1 }}}
  \newcommand{\NOTE}[1]{\phantom{}\begingroup\relax\ifmmode\boldmath\else\bfseries\fi\color{Cerulean}\ignorespaces#1\ignorespaces\endgroup}
  \newcommand{\TODO}[1]{\phantom{}\begingroup\relax\ifmmode\else\sffamily\fi\color{BurntOrange}\ignorespaces#1\ignorespaces\endgroup}
  \newcommand{\FIXME}[1]{\phantom{}\begingroup\relax\ifmmode\boldmath\else\bfseries\sffamily\fi\color{Red}\ignorespaces#1\ignorespaces\endgroup}
  \newcommand{\FIXED}[1]{\phantom{}\begingroup\relax\ifmmode\else\sffamily\fi\color{Green}\ignorespaces#1\ignorespaces\endgroup}
  \newcommand{\REPLACE}[1]{\phantom{}\begingroup\relax\ifmmode\else\sffamily\fi\color{Purple}\ignorespaces#1\ignorespaces\endgroup}
  \newcommand{\DELETE}[1]{\phantom{}\begingroup\relax\ifmmode\else\sffamily\fi\color{Red}\ifmmode\text{\sout{\ensuremath{#1}}}\else\sout{\ignorespaces#1\ignorespaces}\fi\endgroup}
\else
  \newcommand{\davidnote}[1]{}
  \newcommand{\raynote}[1]{}
  \newcommand{\ozgurnote}[1]{}
  \newcommand{\NOTE}[1]{}
  \newcommand{\TODO}[1]{}
  \newcommand{\FIXME}[1]{#1}
  \newcommand{\FIXED}[1]{#1}
  \newcommand{\DELETE}[1]{}
  \newcommand{\REPLACE}[1]{#1}
\fi

\def\BibTeX{{\rm B\kern-.05em{\sc i\kern-.025em b}\kern-.08em
    T\kern-.1667em\lower.7ex\hbox{E}\kern-.125emX}}

\begin{document}

\title{AI Surrogate Model for Distributed Computing Workloads}


\author{
\IEEEauthorblockN{
David K. Park\IEEEauthorrefmark{2},
Yihui Ren\IEEEauthorrefmark{2},
Ozgur O. Kilic\IEEEauthorrefmark{2},
Tatiana Korchuganova\IEEEauthorrefmark{1},
Sairam Sri Vatsavai\IEEEauthorrefmark{2},\\
Joseph Boudreau\IEEEauthorrefmark{1},
Tasnuva Chowdhury\IEEEauthorrefmark{2},
Shengyu Feng\IEEEauthorrefmark{3},
Raees Khan\IEEEauthorrefmark{1}, 
Jaehyung Kim\IEEEauthorrefmark{3},
Scott Klasky\IEEEauthorrefmark{4},\\
Tadashi Maeno\IEEEauthorrefmark{2},
Paul Nilsson\IEEEauthorrefmark{2},
Verena Ingrid Martinez Outschoorn\IEEEauthorrefmark{5},
Norbert Podhorszki\IEEEauthorrefmark{4}, \\
Frederic Suter\IEEEauthorrefmark{4}, 
Wei Yang\IEEEauthorrefmark{6},
Yiming Yang\IEEEauthorrefmark{3},
Shinjae Yoo\IEEEauthorrefmark{2},
Alexei Klimentov\IEEEauthorrefmark{2}, 
Adolfy Hoisie\IEEEauthorrefmark{2}
}
    

    \IEEEauthorblockA{\IEEEauthorrefmark{2}Brookhaven National Laboratory, Upton, NY, USA, \IEEEauthorrefmark{4}Oak Ridge National Laboratory, Oak Ridge, TN, USA}
    \IEEEauthorblockA{\IEEEauthorrefmark{1}University of Pittsburgh, Pittsburgh, PA, USA, \IEEEauthorrefmark{3}Carnegie Mellon University, Pittsburgh, PA, USA} 
    \IEEEauthorblockA{\IEEEauthorrefmark{5}University of Massachusetts, Amherst, MA, USA, \IEEEauthorrefmark{6}SLAC National Accelerator Laboratory, Menlo Park, CA, USA}
}

\maketitle

\begin{abstract}

Large-scale international scientific collaborations, such as ATLAS, Belle II,
CMS, and DUNE, generate vast volumes of data. These experiments necessitate
substantial computational power for varied tasks, including structured data processing,
Monte Carlo simulations, and end-user analysis. Centralized workflow and data
management systems are employed to handle these demands, but current
decision-making processes for data placement and payload allocation are often
heuristic and disjointed. 
This optimization
challenge potentially could be addressed using contemporary machine learning methods, such as
reinforcement learning, which, in turn, require access to extensive data and an
interactive environment. Instead, we propose a generative surrogate modeling
approach to address the lack of training data and concerns about privacy
preservation. We have collected and processed real-world job submission records, 
totaling more than two million jobs through 150 days, and applied four
generative models for tabular data---TVAE, CTAGGAN+, SMOTE, and TabDDPM---to these datasets, thoroughly evaluating their performance. Along with measuring the
discrepancy among feature-wise distributions separately, we also evaluate
pair-wise feature correlations, distance to closest record, and responses to pre-trained
models. Our experiments indicate that SMOTE and TabDDPM can generate similar
tabular data, almost indistinguishable from the ground truth. Yet, as a
non-learning method, SMOTE ranks the lowest in privacy preservation. As a result, we conclude that the probabilistic-diffusion-model-based TabDDPM is the most
suitable generative model for managing job record data.

\end{abstract}




\section{Introduction}
\label{sec:intro}
In a shared high-performance computing (HPC) environment, how to allocate jobs optimally remains a challenging problem. In its most fundamental form, a job scheduling problem bears two-dimension dynamics and is a stochastic knapsack problem~\cite{kleywegt2001dynamic} with an infinite horizon. Each job has at least two dimensions: requested computational resources, such as the number of nodes, and running time, which is unknown to the scheduling system at the time of job submission. As heterogeneous computing architectures become ubiquitous in modern HPC systems, sharing resources, such as central processing unit (CPU), graphics processing unit (GPU), and memory, within a node further complicates the job scheduling problem~\cite{10.1145/3077839.3077855,tang2020cpu}. 
Given an inter-node networking configuration, assigning jobs so their communication does not interfere with each other~\cite{lanOptimizationTopologyAwareJob2023,10.1145/3431379.3460635,10.1109/SC.2018.00029} introduces another optimization dimension. Other research threads focus on reducing the total energy cost by either optimizing under a certain power cap~\cite{10.1145/3547276.3548630} or reducing cooling costs~\cite{mengSimulationOptimizationHPC2015}. 

\begin{figure}
    \centering
    \includegraphics[width=0.95\linewidth]{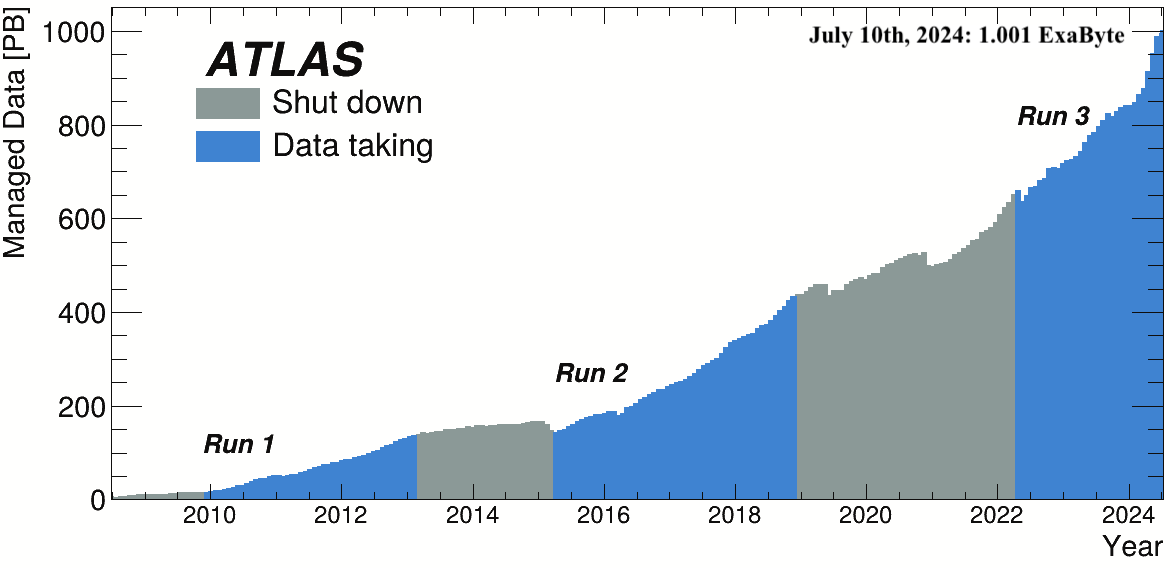}
    \caption{The ATLAS experiment's growing data volume is distributed among computing sites globally.}
    \label{fig:datavol}
    \vspace*{-0.5cm}
\end{figure}

Modern scientific experimental facilities produce large amounts of data at an increasing rate, approaching exabytes. To store, process, and analyze such data, a complex and distributed computing system is required. This leads to a unique shared computing paradigm: high-throughput computing (HTC).
The experimental high energy physics (HEP) community has long benefited from globally distributed computing facilities. For example, the ATLAS collaboration~\cite{atlas, barreiro2017atlas, atlas_dataset_nomenclature} has 182 participating institutions across 42 countries. There are about 150 computing sites, each equipped with different amounts of computing and storage resources. Data accumulated by the ATLAS experiment has reached the exabyte scale (Fig.~\ref{fig:datavol}). Unlike computational fluid dynamics~\cite{10.1016/j.parco.2012.10.002} or \textit{ab initio} molecular dynamics simulations~\cite{10.1016/j.parco.2022.102920} that require intensive computation and communication among worker nodes, the computation workload in experimental physics is often highly parallelizable and input/output (IO) heavy. Scientific discoveries in experimental particle physics are derived from statistics, which then require a large amount of data either by running Monte Carlo simulations or colliding high-energy particle beams repeatedly. The nature of these independent and distributable workloads in particle physics leads to HTC. Therefore, distributed dataset placement and dynamic job allocation play central roles in optimizing HTC systems.

\begin{figure}[t]
    \centering
    \includegraphics[width=0.95\linewidth]{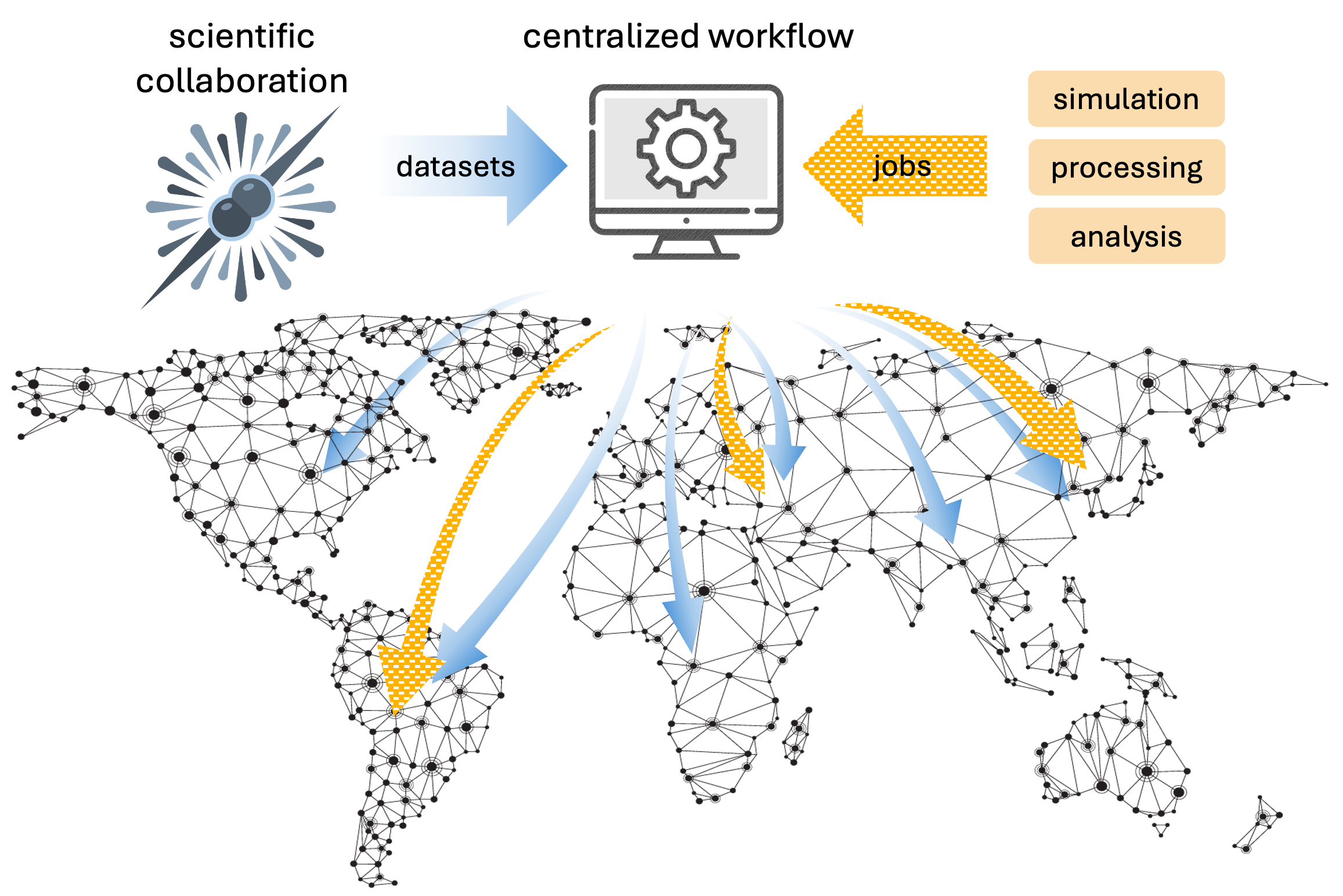}
    \caption{Optimization of data placement and job allocation for distributed computing sites pose challenges to computational resilience and efficiency.
}
    \label{fig:panda}
    \vspace*{-0.3cm}
\end{figure}

Contemporary artificial intelligence (AI)-based optimization algorithms~\cite{van2016deep,schulman2017proximal,mnih2016asynchronous,fujimoto2018addressing,dabney2018implicit,sun2023difusco} require extensive amounts of training data that often are insufficient in real-world applications. Admittedly, on one hand, there are conventional training-free optimization algorithms, such as mixed integer programming~\cite{wolsey2014integer}, evolutionary algorithms~\cite{simon2013evolutionary}, and Bayesian optimization~\cite{shahriari2015taking}, that often suffer from scalability issues and the curse of dimensionality, preventing them from being deployed as a real-time solution. On the other hand, deep learning (DL)-based algorithms are known for their fast inference and flexibility in adapting to different optimization problems. The most widely used DL algorithms for optimization include reinforcement learning~\cite{sutton2018reinforcement} and probabilistic diffusion models~\cite{ho2020denoising,sun2023difusco}. However, such learning-based models require large-scale, high-quality training data and extensive learning periods. Here, we propose to tackle the lack of training data problem with a generative modeling approach that constructs surrogate models to produce synthetic yet realistic workloads.

This work analyzes the data collected from the ATLAS experiment and introduces a generative approach for synthesizing structured tabular job records. As the job records are represented in a tabular format, consisting of mixed categorical and numerical features, conventional neural architectures dedicated for natural images or language processing are not directly employable. We instead rely on recent findings about generative models tailored for structured tabular data. 


\begin{figure*}[t]
    \centering
    \includegraphics[width=1.0\linewidth]{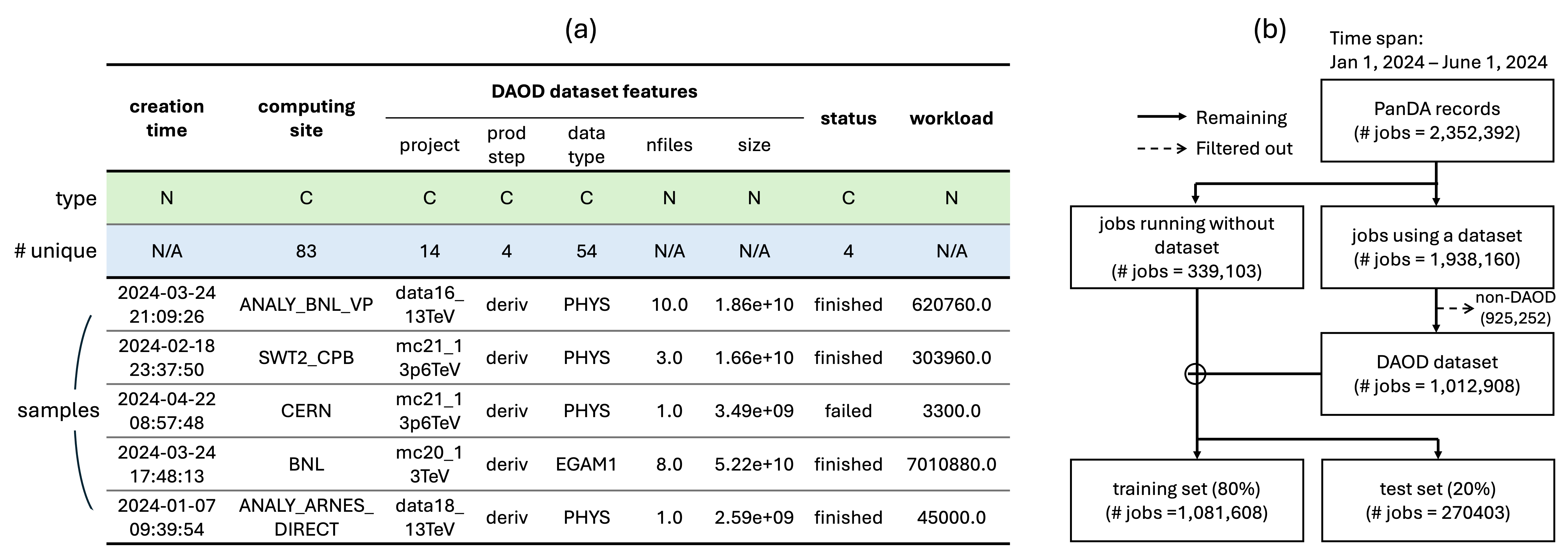}
    \caption{Dataset profile and filtering diagram. \textmd{(a) The feature types (N: numerical; C: categorical) and the number of unique entries (\# unique) reflect the merged training and test data. \emph{creationtime} defines when the job was created. \emph{computingsite} is where the job is executed. Five dataset-related features exist, consisting of the project name (\emph{project}), production step (\emph{prodstep}), dataset type (\emph{datatype}), number of files (\emph{ninputdatafiles}, i.e., \emph{nfiles}), and size of the gross input (\emph{inputfilebytes}, i.e., \emph{size}). The first six features are known prior to running the job, but the latter two features, namely \emph{jobstatus} and \emph{workload}, are unknown until the job is completely executed, defined as the multiplication of number of cores, Gflop per core, and CPU time used. (b) The diagram shows the gross number of PanDA records collected, followed by filtering operations that reduce down to the training and test sets for the generative models.}}
    \label{fig:consort}
    \vspace*{-0.3cm}
\end{figure*}




\section{Data Preparation and Analysis}
\label{sec:dataanalysis}


In this study, we focus on real records of job submission and status in the ATLAS experiment at CERN's Large Hadron Collider (LHC)~\cite{atlas,lhc}. ATLAS is a global scientific collaboration, consisting of over 6000 members. In 2012, the Compact Muon Solenoid (CMS), another scientific collaboration at CERN, discovered the Higgs boson~\cite{higgs_boson}. 
With its globally distributed computing and data storage sites, the ATLAS experiment requires a workflow management system that focuses on data. The Production and Distributed Analysis (PanDA) system~\cite{panda} is engineered to address this by operating at the LHC data processing scale.
Its main purpose is to distribute and manage the large-scale processing of data generated by the experiment. PanDA handles various types of workflows, including user analysis, centralized production, and more complex workflows, ensuring efficient use of distributed computing resources. As centralized production jobs are predictable and well orchestrated, there is not much margin for further improvement. Therefore, this work focuses on user analysis jobs. 

On the storage side, PanDA works in conjunction with Rucio~\cite{barisits2019rucio}, a data management system designed to manage exabyte-scale data volumes in distributed heterogeneous environments~\cite{rucio}. 
Rucio handles the complex requirements of data replication, access, and deletion across multiple storage sites. 
When users submit a computation task or workflow to PanDA, they specify which datasets to process and the software to use. Optionally, the user can provide specific requirements. 
PanDA registers the workflow and divides it into a set of jobs. Then, it chooses the computing resources for executing the jobs based on the availability of input datasets and the workflow's characteristics and requirements (illustrated in Fig.~\ref{fig:panda}).

To provide a general technical understanding and discern the feasibility of using surrogate models for generating workflow features, we have made some simplifications. For example, in the real PanDA system, a user-submitted workflow will be broken down into smaller jobs and briefly verified on a computing site before launching all of them. In this study, we work directly at the job level. Another simplification is to down-select the features from the original PanDA records, which contain more than 100 feature columns. 
As the main goal of this surrogate model is to provide realistic synthetic data for optimizing job and data allocation, we include the job creation time, job status, several dataset-related features, and derive the total computation workload. 


The dataset type interfacing between central production workflows and user analysis workflows is called \textit{derived analysis object data} (DAOD).
DAOD are processed from real experimental data or Monte Carlo simulations, both in a centralized production~\cite{atlas_data_processing_chain}.
As a result, DAOD contribute not only to the largest portion of storage but also the majority of network transmission. 
In this work, we filter out non-DAOD jobs as shown in Fig.~\ref{fig:consort}(b). 
DAOD is registered in the PanDA record as a single entity indicated by its name. However, in our data collection period, most have been used only once or twice. Directly asking the surrogate model to produce a DAOD name is infeasible and makes it difficult to validate and compare the models. 
Thankfully, DAOD names consist of several meaningful sections, such as \emph{project}, production step ( or \emph{prodstep} for short), and \emph{datatype}~\cite{atlas_dataset_nomenclature} (refer to DAOD dataset features in Fig.~\ref{fig:consort}(a)). 
We separate the field of DAOD names into these categorical features along with the total number of input files (\emph{ninputdatafiles}) and \emph{inputfilebytes} for each job.

To compute the job workload, we extract the job running time and number of cores used from the PanDA records and scale the core-hours by the processing power of the assigned computing site. 
The processing power of a computing site is obtained from the high energy physics computing score benchmark (HS23)~\cite{hs23} based on a suite of real-world data processing and simulation tasks in HEP.

\section{Related Works}
\label{sec:related}

In job scheduling scenarios involving large, complex, and heterogeneous computing environments, heuristic algorithms typically experience significant performance drops~\cite{noheu}, while neural-network-based algorithms~\cite{jobscheduling1, jobscheduling2, jobscheduling3} tend to maintain more efficient scheduling. Reinforcement learning is a popular model choice for job scheduling applied on the optimization of device placement~\cite{jobscheduling2}, distributed computing~\cite{jobscheduling3}, or data-parallel cluster scheduling~\cite{jobscheduling1}. However, ensuring safety during both the training and deployment phases poses a significant challenge with reinforcement learning~\cite{gu2022review}. Reinforcement learning often has difficulty observing the proper balance between safety and task performance, producing policies that are either too risky or overly cautious. This issue is especially critical in applications where unsafe behaviors can result in catastrophic consequences~\cite{xu2022trustworthy}, such as within the ATLAS collaboration. For a safe application of reinforcement learning methods, specialized benchmark models and datasets to enable offline safe learning have been emphasized~\cite{liu2023datasets}. Extending this effort, we attempt to generate novel synthetic data for PanDA records that affords a reduced risk in the optimization of distributed computing.

Tabular data are defined as structured tables consisting of both categorical and numerical features. Tabular datasets typically are limited in size, in contrast to popular vision or natural language processing problems that benefit from abundant data readily available on the Internet. Generative models for tabular data are actively investigated in the machine learning community due to significant demand for high-quality synthetic data for tabular data augmentation. Responding to these needs, tabular models have been developed based on deep generative models, leading to competitive performances~\cite{xu2019modeling, engelmann2021conditional, fan2020relational, jordon2018pate, kim2022sos, torfi2022differentially, zhang2021ganblr, zhao2021ctab}, and the number of papers about tabular generation is exponentially growing. Popular neural architectures include autoencoders (AE)~\cite{ref_autoencoder1, xu2019modeling}, generative adversarial networks (GANs)~\cite{xu2019modeling, zhao2021ctab, zhao2024ctab}, transformers~\cite{ref_transformer1, ref_transformer2}, or diffusion models~\cite{ref_diffusion1, kotelnikov2023tabddpm}.

Prior studies employ several publicly available datasets, such as OpenML~\cite{OpenML2013}, for objective evaluations. Data size of the frequently used public datasets may be on the order of between 100 and 100,000~\cite{kotelnikov2023tabddpm}. These datasets also have a designated target feature for either classification or regression~\cite{ref_diffusion1} with or without a timestamp column to show time-dependent variations~\cite{kotelnikov2023tabddpm}. PanDA records differ from existing public datasets in several areas. First, the size of the PanDA records is extensive, reaching more than two million rows for 150 days. Second, the records are complex and heterogeneous, containing columns of multiple users, computing sites, and physics datasets, with the counts often imbalanced. Third, the number of job submission records fluctuates over time, showing clear time-varying patterns in distribution. These characteristics pose distinct challenges in generative model training for PanDA records.

\section{Method: Generating Surrogate Models}
\label{sec:method}


\begin{figure*}[t]
    \centering
    \includegraphics[width=1\linewidth]{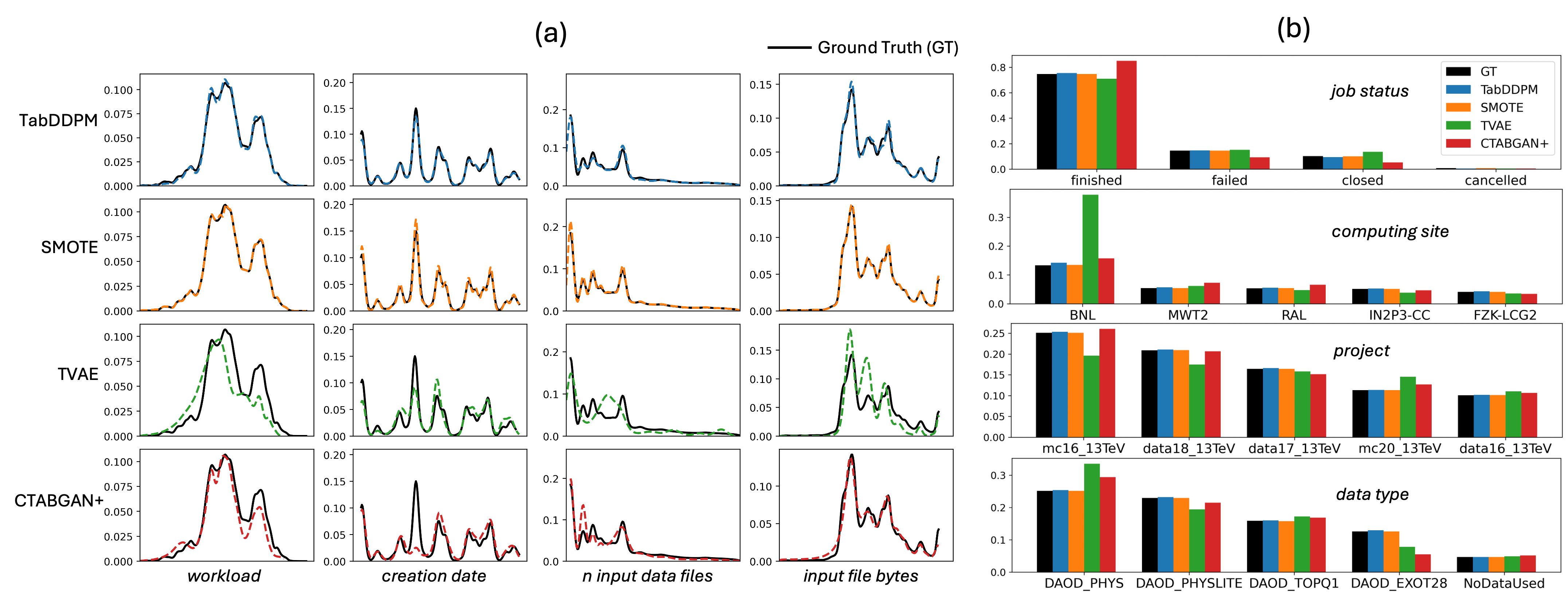}
    \caption{Comparisons of generative performances  based on distributional similarities of individual features. \textmd{(a) Distinct columns show each of all four numerical features used as training inputs, while individual rows correspond to a model. Black and dotted color lines correspond to ground truth (GT) and synthetic data, respectively. (b) The graphs are comparing if distributions are similar for unique entries with top counts across four categorical features.}}
    \label{fig:distribution}
    \vspace*{-0.3cm}
\end{figure*}



\subsection{Generative Surrogate Models}

We consider four baselines for the generation of synthetic PanDA records. 
First, TVAE~\cite{xu2019modeling} uses a variational autoencoder (VAE)~\cite{vae} as the backbone for learning and synthesizing mixed-type tabular data. VAE is composed of an encoder that encodes each row of the training data as a latent code, and a decoder that reconstructs the input data from the latent code. During the synthesis stage, latent codes are sampled, followed by a forward propagation to the decoder for generating novel data. VAE is trained by minimizing the reconstruction error and the KL (Kullback–Leibler) divergence loss between the latent code and Gaussian distribution.

Second, CTABGAN+~\cite{zhao2024ctab} is the current state of the art among tabular generation models based on GANs~\cite{gan}. A GAN is composed of two neural network architectures, namely a generator and a discriminator. The discriminator's objective is to distinguish whether the given input is real or synthetic data, while the generator tries to synthesize new data that can trick the discriminator, optimally converging to a point where the discriminator cannot distinguish the synthesized data. CTABGAN+ adapts the GAN to accept mixed continuous and categorical features while improving generation quality compared to its former variant CTABGAN~\cite{zhao2021ctab}.

Meanwhile, SMOTE~\cite{chawla2002smote} is the only baseline model that runs without learning. It was originally introduced to address imbalanced datasets via oversampling minority classes, synthesizing new data using a nearest-neighborhood method. Albeit simple, SMOTE has demonstrated competitive performances even compared to recently introduced models based on neural networks. 

TabDDPM~\cite{kotelnikov2023tabddpm} employs diffusion models for modeling tabular data, achieving competitive results in tabular generation. Diffusion models~\cite{ho2020denoising} are a type of generative model used originally for image synthesis. The core idea is to begin with random noise and iteratively refine it to produce structured data, such as an image or tabular data. Diffusion models follow two sequential processes for synthesis, namely a forward (i.e., diffusion) and backward (i.e., denoising) process. In the forward process, the model gradually adds noise to a data point (e.g., a row of tabular data) over multiple steps, making it increasingly random until it becomes pure noise. The model then learns to denoise the data in the reverse process, starting from noise and progressively removing the added noise to recover the original tabular data. This reverse process generates new row data that resembles the training data. TabDDPM employs a latent diffusion model~\cite{rombach2022high} as the model backbone for generation, while using multi-layer perceptrons (MLPs) within encoding and decoding layers.

\subsection{Evaluation Metrics}

\paragraph{Per-feature evaluation} One simple way to evaluate generative performance is by measuring the distribution of individual features of the real data followed by quantifying the similarity of each feature to synthetic counterparts. To measure the divergence, we compute the Wasserstein distance (WD) between numerical features and the Jensen–Shannon divergence (JSD) between categorical ones, following the convention in~\cite{zhao2021ctab, kotelnikov2023tabddpm}. These metrics measure whether each column of the synthetic data follows the distribution of the real data. A small WD and JSD denote that on average, the generative model is capable of producing realistic synthetic data per feature. However, WD and JSD cannot measure if the model also learns covariance or joint distribution of different input features.

\paragraph{Correlations between feature pairs} For realistic synthesis, a tabular generative model must learn the correlated structures of the incorporated features properly. Hence, we also report pair-wise correlations to demonstrate the ability to learn the covariance. We plot the lower triangle of the correlation matrix for the ground truth training data in Fig.~\ref{fig:correlation}(a). Pearson correlation, correlation ratio, and Theil's U statistic are used to measure the correlation between a pair of numerical features, numerical-categorical features, and categorical features, respectively. If the generative model precisely learns the correlated structure between columns, the correlation matrices of real and synthetic data should be similar element-wise. This performance is evaluated by a mean L2 distance between real and synthetic correlation matrices (denoted as ``diff-CORR'').

\paragraph{Measuring fully joint distribution via MLEF} Lastly, a model's capacity in learning the true joint distribution of the training data is evaluated via machine learning efficacy (MLEF)~\cite{xu2019modeling, zhao2021ctab, kotelnikov2023tabddpm}. MLEF records whether synthesized tabular data can be used as training data for predicting a target feature in the test data. In this work, the numeric feature of \textit{workload} is used as the target feature to predict, and a mean-squared error is used as the performance measurement for the regression. In practice, CatBoost~\cite{prokhorenkova2018catboost} is used as the regressor for the workload prediction task, where the target feature is transformed with a natural-log to avoid scale-dependent instability during training. A smaller MLEF indicates that synthesized data consist of information related to predicting the workload, and the generative model learns the relation successfully. We report diff-MLEF, which shows the difference of the synthetic data MLEF versus the real training data ($\text{diff-MLEF} \coloneq \text{MLEF}_{\text{synthetic}} - \text{MLEF}_{\text{train}}$). A small diff-MLEF is desirable but with a theoretical minimum of zero at which point the synthesized data have equal value as the ground truth in the workload prediction.

\paragraph{Measuring privacy preservation via DCR} While the aforementioned metrics evaluate the capacity to learn true distribution of the training data, these metrics fall short of detecting if the training data are simply memorized and repeated during synthesis. Avoiding such trivial memorization is important in terms of privacy concerns. According to regulations such as General Data Protection Regulation (EU), California Consumer and New York Privacy Acts (US), and General Data Protection Law (LGPD, Brazil), synthetic datasets should not include real user data, so they can be shared publicly without compromising anonymity. We measure the Distance to Closest Record (DCR) to assess privacy risk of synthetic data. For computing DCR, a single record in the training data closest to a synthetic instance is identified, and the distance is averaged over all synthetic data. A small DCR indicates the synthetic data closely follow the original data instances, showing the model merely mimics the training data while posing a greater privacy risk.

\vspace{-0.2cm}


\section{Experiments}
\label{sec:experiments}

\subsection{Training Details}
\paragraph{Training tabular generative models}
We identify five categorical features --- job status, computing site, project name, production step, data type --- and four numerical features --- workload, creation date, number of input data files, and input file gross byte size --- for the training. Fig.~\ref{fig:consort}(a) shows details of each feature. Numerical and categorical columns are pre-processed separately. Numerical features are normalized via Gaussian quantile transformation from the scikit-learn library~\cite{pedregosa2011scikit}. All individual entries in the categorical columns are regarded unique and represented as a one-hot vector. The 150-day PanDA job records are split into training and test set by 80\% and 20\%, respectively. In total, each model is trained on the training set consisting of 1,319,007 job records. Other hyperparamters are inherited per the experiments in the original papers~\cite{xu2019modeling, zhao2024ctab, kotelnikov2023tabddpm}. Each baseline model is trained for 30,000 epochs with a learning rate of 0.0002, which decays following a cosine scheduler.

\paragraph{Training CatBoost regressor for computing MLEF} CatBoost regressors are trained on five different data separately, including ground truth training data and four synthetic datasets obtained from the trained surrogate models. Each training for CatBoost lasts for 200 iterations with a depth of 10 and a learning rate of 1.0 on root mean square error loss. Each of the trained CatBoost is then evaluated on the test data (Fig.~\ref{fig:consort}(b)).

\begin{figure*}
    \centering
    \includegraphics[width=1\linewidth]{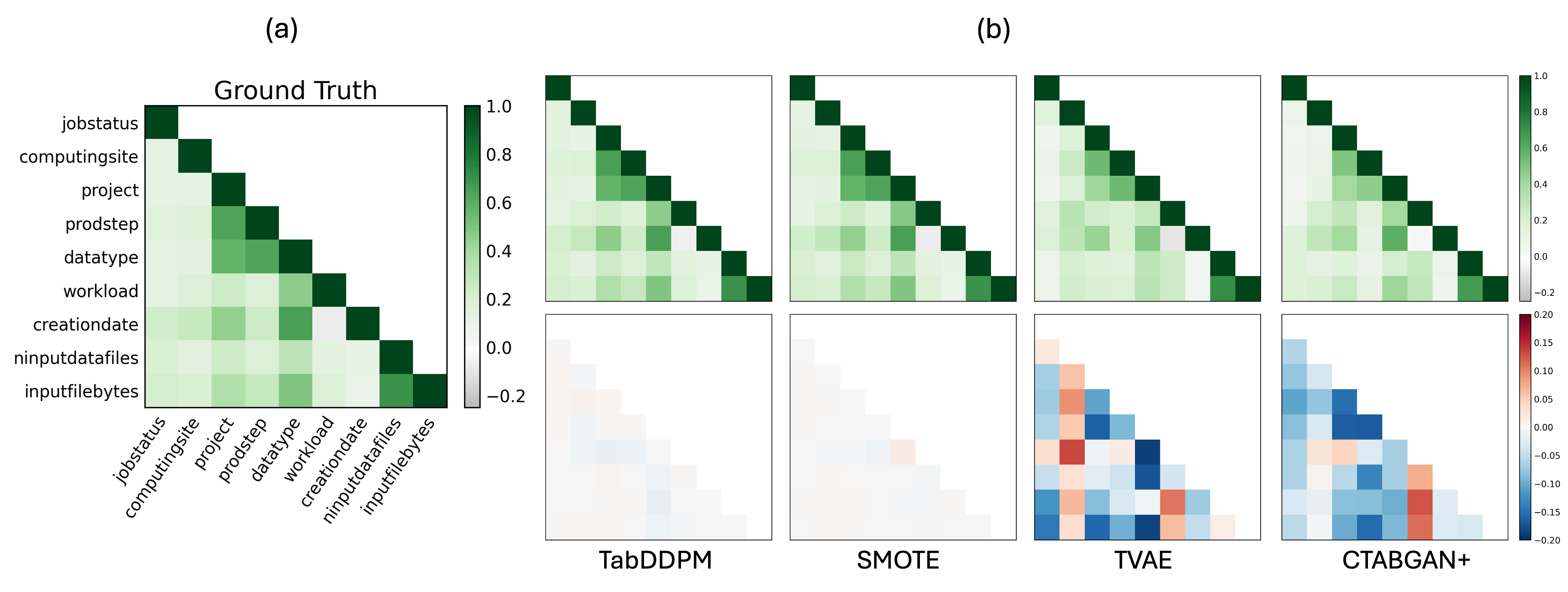}
    \caption{Correlations between features in tabular data. \textmd{(a) Correlation strengths in ground truth training data are shown. (b) Synthetic data correlations are compared across implemented models on tabular generative models. The bottom row shows the difference versus the ground truth.}}
    \label{fig:correlation}
    \vspace*{-0.3cm}
\end{figure*}

\subsection{Results}

\paragraph{Per-feature evaluation}
Fig.~\ref{fig:distribution} depicts the distribution of individual features across ground truth and the baseline models. In Fig.~\ref{fig:distribution}(a), \emph{workload} displays several peaks, which TVAE and CTABGAN+ fail to model accurately, while TabDDPM and SMOTE significantly overlap with the ground truth. The time-varying fluctuation in the number of jobs (\emph{creationdate}) is also shown. Again, TabDDPM and SMOTE learn the temporal distribution successfully, while the other models fail to capture the peaks. In Fig.~\ref{fig:distribution}(b) on four categorical features, normalized count of entries with the top five counts are shown (except for \emph{jobstatus}, which has four). Note that TabDDPM and SMOTE perform equivalently well with the count similar to the ground truth. TVAE amplifies the count for BNL of \emph{computingsite} and DAOD\_PHYS in \emph{datatype}. Quantitative values in Table~\ref{table:surrogate_evaluation} support this finding, where SMOTE and TabDDPM perform competitively on WD and JSD close to each other.

\paragraph{Correlations between feature pairs}
While TVAE and CTABGAN+ perform poorer WD and JSD than SMOTE, the gap is not striking compared to diff-CORR. In Table~\ref{table:surrogate_evaluation}, diff-CORR for SMOTE and TabDDPM are 0.011 and 0.036, respectively, compared to the other two models that exceed 0.65. This difference is also conspicuous in Fig.~\ref{fig:correlation}(b). While all models seem to agree with the ground truth pattern in Fig.~\ref{fig:correlation}(a) on the upper row, the difference with the ground truth on the bottom row reveals that TVAE and CTABGAN+ show a large error on multiple features as the dark red or blue squares indicate.

\begin{table}[]
\begin{center}
\caption{Performance comparisons on surrogate models}
\label{table:surrogate_evaluation}
\begin{tabular}{@{}llllll@{}}
\toprule
Model & WD $\downarrow$    & JSD $\downarrow$    & \begin{tabular}[c]{@{}l@{}}diff-\\ CORR\end{tabular}$\downarrow$  & DCR $\uparrow$   & \begin{tabular}[c]{@{}l@{}}diff-\\ MLEF\end{tabular}$\downarrow$ \\ \midrule
TVAE     & 0.961  & 0.806 & 0.653     & \textbf{0.143} & 5.875\\
CTABGAN+ & 1.0 & 0.820 & 0.658     & \underline{0.105} & 10.464\\
SMOTE     & \textbf{0.871} & \textbf{0.799} & \textbf{0.011}     & 0.001& \textbf{0.058}\\
TabDDPM  & \underline{0.874} & \textbf{0.799} & \underline{0.036}     & 0.025 & \underline{0.826}\\ 
\bottomrule
\end{tabular}
\end{center}
\vspace*{-0.5cm}
\end{table}


\paragraph{diff-MLEF and DCR}
MLEF of SMOTE outperforms other models by a large margin, while recording a low DCR. Because SMOTE is a non-learning algorithm, essentially generating new samples by mixing the five nearest neighbors in the latent space, generated samples tend to bear similarity with the original training data. The low DCR means there is a privacy risk in generating samples that may expose the training data. TabDDPM comparably achieves a higher DCR, relatively free from the privacy risk. Meanwhile, TVAE and CTABGAN+ show much higher DCR, demonstrating a lower risk for breaching privacy. However, DCR can also be elevated when the generative performances are poor, such as when the model simply cannot learn the true distribution. Therefore, all five metrics should be carefully reviewed based on the target goal. If the cost of privacy risk is substantial, SMOTE is not preferable.

\vspace{-0.2cm}

\section{Conclusion}
\label{sec:conclusion}
This study tests whether a PanDA dataset can be synthesized to improve job scheduling optimization, which may contribute to the resiliency of distributed computing. We have collected 150 days of PanDA job records data from the ATLAS collaboration and down-selected core features, such as dataset types and input file sizes, to derive new features, e.g., computation workload, of each job. We have investigated four representative generative models, TVAE, CTAGGAN+, SMOTE, and TabDDPM, as surrogate models, while their respective performance has been thoroughly studied and compared.
SMOTE and TabDDPM record outstanding performances in matching the distributions with the ground truth at the feature-wise level, while TVAE did not perform well on \emph{ninputdatafiles}, \emph{inputfilebytes}, and \emph{computingsites}.
In terms of pairwise feature correlations, both TabDDPM and SMOTE resemble such correlations from the real data. 
Synthetic data generated from TabDDPM and SMOTE also respond to a pre-trained supervised model similarly compared to the real data.
However, due to SMOTE's non-learning nature, most synthesized data are too similar to the original and lack of privacy-preserving functionality. 
As such, we expect synthesized data from SMOTE will not provide much value as those from TabDDPM for training an AI-based optimizer or a downstream predictive model.
Overall, TabDDPM strikes a good balance between faithfulness, drawing data from the same distribution, and diversity, recognizing synthetic data differ from real data. 
Such a synthetic data generator will be important for training AI-based optimization algorithms by assigning jobs and allocating data. It also will provide more realistic workload inputs to calibrate large-scale event-based simulations.

Still, we recognize this study has several limitations that may be addressed in near future. First, we assume the dataset is in a tabular form and treat each row, a job, independently. The temporal aspect of the submitted jobs has not been studied in depth. For example, whether or not there are periodic ups and downs due to weekends has not been investigated. Based on the preliminary results of \emph{creationdate} distributions, we maintain these deep generative models can reproduce periodic temporal patterns. 
Second, we presume the majority of the jobs are normal operations, and the distributed computing systems perform normally. However, it is unclear if such a generative modeling approach can be extended to abnormal scenarios. From past experience in applying diffusion models for particle physics data as a surrogate model~\cite{go2024effectiveness}, the data scarcity region usually exhibits a higher error rate. Interestingly, this characteristic of diffusion models makes it a competent detector for anomalies~\cite{livernoche2024on}. 
Third, data collection and processing can be further improved. The duration of the data collection could be extended to years at the cost of potentially more complicated procedures to manage different data formats and structures. Similar work might be done from the dataset perspective to predict dataset reuse factors or identify popular datasets.

\section*{Acknowledgments}

This material is based on work supported by the U.S. Department of Energy, Office of Science, Office of Advanced Scientific Computing Research under Award Number DE-SC-0012704. This work was done in collaboration with the distributed computing research and development program within the ATLAS Collaboration. We thank our ATLAS colleagues for their support, particularly the ATLAS Distributed Computing team's contributions. We would also like to express our deepest gratitude to Prof.~Kaushik De at the University of Texas at Arlington.

\vspace{25pt}

\vspace{-5pt}
\clearpage
\bibliographystyle{IEEEtran}
\bibliography{ref.bib}

\end{document}